\documentclass[a4paper,12pt]{article}
\usepackage{epsfig,float}
\usepackage[left=2cm,right=2cm]{geometry}
\usepackage{a4wide,graphicx}

\newcommand{\eps}{\epsilon}
\begin{document}

\title{The $\rho\rho$ interaction in the hidden gauge formalism and the
$f_0(1370)$ and $f_2(1270)$ resonances.}

\author{R. Molina$^1$, D. Nicmorus$^2$ and E. Oset$^1$}
\maketitle

\begin{center}
$^1$ Departamento de F\'{\i}sica Te\'orica and IFIC,
Centro Mixto Universidad de Valencia-CSIC,
Institutos de Investigaci\'on de Paterna, Aptdo. 22085, 46071 Valencia, Spain\\
$^2$ Fachbereich Theoretische Physik, Institut f\"ur Physik,Karl-Franzens-Universit\"at Graz, Universit\"atsplatz 5, A-8010 Graz, Austria\\
\end{center}

\date{}

\small{PACS numbers: 13.75.Lb, 14.40.Cs, 12.40.Vv, 12.40.Yx}

 \begin{abstract}  

We have studied the interaction of vectors mesons within the hidden gauge
formalism and applied it to the  particular case of the $\rho \rho$
interaction. We find a strong attraction in the isospin, spin channels I,S=0,0
and 0,2, which is enough to bind the $\rho \rho$ system. We also find that the
attraction in the I,S=0,2 channel is much stronger than in the 0,0 case. The
states develop a width when the $\rho$ mass distribution is considered, and
particularly when the $\pi \pi$ decay channel is turned on. Using a
regularization scheme with cut offs of natural size, we obtain results in fair
agreement with the mass and the width of the $f_0(1370)$ and $f_2(1270)$ meson
states, providing a natural explanation of why the tensor state is more bound
than the scalar and offering a new picture for these states, which would be
dynamically generated from the $\rho \rho$ interaction or, in simpler words, 
$\rho \rho$ molecular states.

\end{abstract}

\section{Introduction}

Chiral perturbation theory, with its unitary extensions to higher energies has
brought a new momentum to hadron physics at low and intermediate energies. The
exact unitarity in coupled channels together with dispersion relations
\cite{nsd,ollerulf}, the inverse amplitude method (IAM)
\cite{Dobado:1996ps,ramonet} or the equivalent solution in terms of
coupled Bethe Salpeter equations 
\cite{weise,npa,norbert} introduce a nonperturbative 
scheme that proves highly efficient to study meson-meson or meson-baryon
interactions, usually referred to as the chiral unitary approach. By fixing a
minimum of subtraction constants or cut offs to regularize the loops, one
finds a fair agreement with data in a vast amount of reactions 
\cite{review} (see \cite{aligarh} for a recent review). One of the
results of these studies is that the amplitudes have sometimes poles that can be
associated to known resonances. Sometimes new resonances are predicted, like a
second $\Lambda(1405)$ \cite{jidocola} or a second $K_1(1270)$ axial vector
meson \cite{luisaxial}, for which experimental support has been found in 
\cite{magastwolamb} and \cite{gengkk} respectively. So far, resonances have
been investigated in the interaction of the SU(3) octet of the pseudoscalar 
mesons of the $\pi$ with themselves 
\cite{npa,norbert,Markushin:2000fa,Dobado:1996ps,ramonet}, 
which provide the low lying scalar mesons,
 the interaction of the pseudoscalar mesons with
 the octet of baryons of the $p$, which generate $J^P=1/2^-$ baryonic
 resonances
 \cite{weise,angels,ollerulf,carmina,jidocola,carmenjuan,hyodo}, 
 the interaction of pseudoscalar mesons with the
decuplet of the $\Delta$ \cite{kolodecu,sarkar}, which leads to $J^P=3/2^-$ baryon resonances, 
and the interaction of pseudoscalar mesons with vector mesons, which leads to 
axial vector meson resonances \cite{lutzaxial,luisaxial}. Yet, the interaction of
vector mesons with themselves has not been tackled from this perspective. The
purpose of the present paper is to study this interaction and show how, also in
this case, one obtains dynamically generated resonances. 

    The interaction of vector mesons with themselves is done using the
Lagrangians of hidden gauge formalism, which mix vector mesons with
pseudoscalars and respect chiral symmetry \cite{hidden1,hidden2}. The hidden gauge Lagrangians for vector-vector interaction do not provide local chiral Lagrangians 
as in the case of meson-meson or meson-baryon interaction discussed above. 
Non local terms corresponding to the exchange of vector mesons 
appear in the amplitudes. Yet, under certain approximations these terms can also be recast in the form of local Lagrangians similar to those quoted above. In this first
paper on the issue we shall describe the formalism and apply it to study the
$\rho -\rho$ interaction. We shall see that one gets attraction in the $I=0,S=0$
and $I=0,S=2$ channels which is enough to produce bound states of   $\rho -\rho$. We
shall see that the interaction in the $I=0,S=2$ tensor case is stronger than in
the scalar one $I=0,S=0$ and that the states that we obtain can be associated
to the known resonances $f_0(1370)$ and $f_2(1270)$. In order to obtain the
width of the states we shall also consider their decay into $\pi \pi$, obtained
within the same formalism of \cite{hidden1,hidden2}. 

On the theoretical side there is work done for both resonances. The coupling of 
the tensor resonance to $\pi \pi$ was exploited in \cite{fdos}, within the
formalism of the IAM, and the $f_2(1270)$ was obtained qualitatively, at the
expense of adding counterterms or higher order that produce the resonance within
this formalism.  This does not mean that the $f_2(1270)$ is a resonance built up
from $\pi \pi$. The information on the resonance properties is essentially
buried in the counterterms, like in the case of the $\rho$ meson that is also 
obtained within the IAM \cite{Dobado:1996ps,ramonet} starting from the  
$\pi \pi$ interaction. There again, the basic properties of the rho are tied to
the $L_i$ coefficients of the second order chiral Lagrangian 
\cite{Gasser:1984gg}, and its generation within the IAM does not mean that one
has a dynamically generated resonance from the basic $\pi \pi$ interaction.
Indeed, 
a careful study of the large $N_c$ behavior of the resonance shows that the
state remains as  $N_c$ goes to infinite, as corresponds to genuine states that
are built essentially from $q \bar{q}$, unlike the dynamically generated scalar
mesons that fade away in that limit \cite{largenc}. In the present case
the counterterms needed in the IAM to produce this state in \cite{fdos} are
burying the information about the nature of the $f_2(1270)$ resonance, which, as
we will show, gets dynamically generated from the $\rho\rho$ interaction. Work
with quark models is also available. In \cite{Fariborz:2006xq} the $f_0(1370)$
is assumed to be dominantly a $q \bar{q}$ state, unlike the lighter scalars
that are assumed to be largely four quark states. In \cite{Umekawa:2004js} the
$f_0(1370)$ is also studied within the improved ladder approximation of QCD
assuming it to be mostly made of $q \bar{q}$ components, although, as quoted
there, the meson-meson or four quark components are supposed to be important. 
In \cite{Rodriguez:2004tn} the $f_0(1370)$ is assumed to be a mixture of 
$q \bar{q}$ and four quarks, while in  \cite{Close:1996sv} a mixture of 
$q \bar{q}$ components with glueballs is preferred. Once again, in 
\cite{Kleefeld:2001ds} the $q \bar{q}$ nature is preferred for the $f_0(1370)$
with the quarks of non strange nature. For the case of the $f_2(1270)$ there is
also work done in \cite{Anisovich:2001zp}, where the state is assumed to be
predominantly a $q \bar{q}$ state.

   Our work will bring a new perspective into this panorama, showing that 
practically with no freedom (up to fine tuning of a cut off parameter from
values around the natural size), the $f_2(1270)$ and $f_0(1370)$ states emerge
as bound states of the $\rho \rho $ interaction evaluated within the reliable
formalism of hidden gauge.

\section{Formalism for $VV$ interaction}

We follow the formalism of the hidden gauge interaction for vector mesons of 
\cite{hidden1,hidden2}(see also \cite{hidekoroca} for a practical set of Feynman rules). 
The interaction Lagrangian involving the interaction of 
vector mesons amongst themselves is given by
\begin{equation}
{\cal L}_{III}=-\frac{1}{4}\langle V_{\mu \nu}V^{\mu\nu}\rangle \ ,
\label{lVV}
\end{equation}
where the symbol $\langle \rangle$ stands for the trace in the $SU(3)$ space 
and $V_{\mu\nu}$ is given by 
\begin{equation}
V_{\mu\nu}=\partial_{\mu} V_\nu -\partial_\nu V_\mu -ig[V_\mu,V_\nu]\ ,
\label{Vmunu}
\end{equation}
with $g$ given by
\begin{equation}
g=\frac{M_V}{2f}\ ,
\label{g}
\end{equation}
with $f=93\,MeV$ the pion decay constant. The value of $g$ of eq. (\ref{g}) is 
one of the ways to account for the $KSFR$ rule \cite{KSFR}, which is tied to 
vector meson dominance \cite{sakurai}. The magnitude $V_\mu$ is the $SU(3)$ 
matrix of the vectors of the octet of the $\rho$
\begin{equation}
V_\mu=\left(
\begin{array}{ccc}
\frac{\rho^0}{\sqrt{2}}+\frac{\omega}{\sqrt{2}}&\rho^+& K^{*+}\\
\rho^-& -\frac{\rho^0}{\sqrt{2}}+\frac{\omega}{\sqrt{2}}&K^{*0}\\
K^{*-}& \bar{K}^{*0}&\phi\\
\end{array}
\right)_\mu \ .
\label{Vmu}
\end{equation}

\begin{figure}
\begin{center}
\includegraphics[width=16cm]{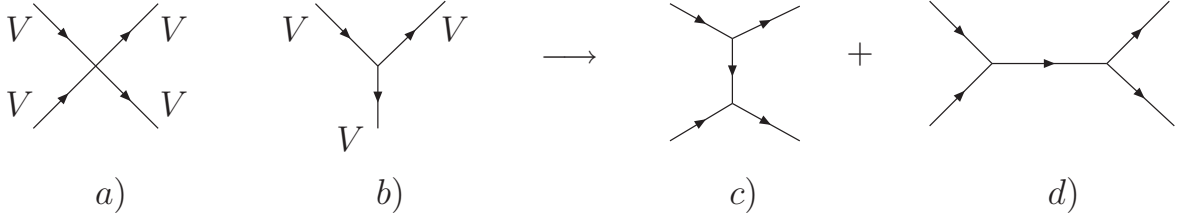}
\end{center}
\caption{Terms of the ${\cal L}_{III}$ Lagrangian: a) four vector contact term,
 eq. (\ref{lcont}); b) three vector interaction, eq. (\ref{l3V}); c) $t$ and 
 $u$ channels from vector exchange; d) $s$ channel for vector exchange.}
\label{fig:fig1} 
\end{figure}

The interaction of ${\cal L}_{III}$ gives rise to a contact term coming for 
$[V_\mu,V_\nu][V_\mu,V_\nu]$
\begin{equation}
{\cal L}^{(c)}_{III}=\frac{g^2}{2}\langle V_\mu V_\nu V^\mu V^\nu-V_\nu V_\mu
V^\mu V^\nu\rangle\ ,
\label{lcont}
\end{equation}
depicted in fig. \ref{fig:fig1} a), and on the other hand it gives rise to a three 
vector vertex
\begin{equation}
{\cal L}^{(3V)}_{III}=ig\langle (\partial_\mu V_\nu -\partial_\nu V_\mu) V^\mu V^\nu\rangle
\label{l3V}\ ,
\end{equation}
depicted in fig. \ref{fig:fig1} b). This latter Lagrangian gives rise to a
$VV\to VV$ interaction by means of the exchange of one of the vectors, as 
shown in fig. \ref{fig:fig1} c). These Lagrangians have been previously used to study collision rates of vector mesons in heavy ion collisions \cite{luisruso}.

The $SU(3)$ structure of the Lagrangian allows us to take into account all the
channels within $SU(3)$ which couple to certain quantum numbers. This is what
is done in the study of the interaction of pseudoscalar mesons
\cite{npa,norbert} where, for instance, for the scalar-isoscalar channel one
introduces the $\pi\pi$ and $K\bar{K}$ pairs as coupled channels. In this
particular case the interaction leads to the generation of two scalar isoscalar
resonances, the $f_0(600)$ or $\sigma$ and the $f_0(980)$. It is also seen that
the $\pi\pi$ and $K\bar{K}$ states largely decouple: the $\sigma$ is basically
 $\pi\pi$ resonance, while the $f_0(980)$ couples mostly to $K\bar{K}$ and
weakly to $\pi\pi$, as a consequence of which one has a small width for the
$f_0(980)$ in spite of the large phase space for decay into $\pi\pi$.

In the present work we shall provide the formalism for the $VV$ interaction and
present results for the simplest case, the $\rho\rho$ interaction. We are
aiming at obtaining from this interaction the lightest scalar and
tensor mesons after the $f_0(980)$ which is very well reproduced in terms of
pseudoscalar-pseudoscalar components. In the PDG \cite{pdg} we find the
$f_2(1270)$ and the $f_0(1370)$. The $\rho\rho$ system interacting in $S$ wave,
as we shall do, can lead to different isospin, spin states
$I,J=0,0;\,1,1;\,0,2;\,2,0$ and $2,2$. It would be most interesting to see if
the results that we obtain agree, at least qualitatively, with the experimental
data on this sector at low energies. In particular, it is challenging to find a
reason on why the $f_2(1270)$ is lower in energy than the $f_0(1370)$. In this
exploratory work we shall work only with $\rho$ meson, with the limited aim to
learn about the structure of these low lying resonances. The use of couple
channels is welcome and should be tackled in the future, but for purpose of
studying these low lying resonances, the other channel, $K^*\bar{K}^*$, has a
mass of $1784\,MeV$, more than $400\,MeV$ higher than the mass of the
$f_0(1370)$, and then, as is the case in all studies of meson-meson
interaction, this channel can barely affect the structure of these low lying
resonances. In any case, its possible effect, through a weakly energy dependent
$K\bar{K}$ loop function at the energies that we are concerned, can be
accommodated by fine tuning the subtraction constants of the regularized
$\rho\rho$ loop function, or equivalently the cut off, as we shall do. 

\begin{figure}
\begin{center}
\includegraphics{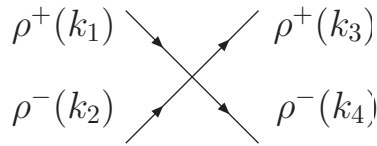}
\end{center}
\caption{Contact term of the $\rho\rho$ interaction.}
\label{fig:fig2}
\end{figure}

Starting with the Lagrangian of eq. (\ref{lcont}) we can immediately obtain the 
corresponding amplitude to $\rho^+\rho^-\to \rho^+\rho^-$ corresponding to 
fig. \ref{fig:fig2}.

We immediately obtain\footnote{We always use a carthesian basis for the polarization vectors. Should one use a spherical basis, one should complex conjugate the $\eps_{\mu}$ of the outgoing vectors.}:
\begin{eqnarray}
-it^{(c)}_{\rho^+\rho^-\to\rho^+\rho^-}= i2g^2 (2\eps^{(1)}_\mu \eps^{(2)}_\nu \eps^{(3)}_\nu \eps^{(4)}_\mu -\eps^{(1)}_\mu \eps^{(2)}_\mu \eps^{(3)}_\nu \eps^{(4)}_\nu -\eps^{(1)}_\mu \eps^{(2)}_\nu \eps^{(3)}_\mu \eps^{(4)}_\nu )\ ,
\label{rhorho}
\end{eqnarray}
where the indices $1,2,3$ and $4$ correspond to the particles with momenta 
$k_1,k_2,k_3$ and $k_4$ in fig. \ref{fig:fig2}. For simplicity of the notation
we write the Lorentz indices as subindices with the understanding that repeated
indices should be one covariant and the other one contravariant.

Eq. (\ref{rhorho}) shows three different structures of the vector
polarizations, the same number as possible spins of the two $\rho$ systems, and
there is some relationship as we shall see below.

The large mass of the vectors offers a technical advantage, since the
three momenta of the $\rho$ in the scattering amplitude in the region of energies of
interest to us are small compared to its mass. We shall work in the limit of
small three momenta of the $\rho$ where the $\eps_\mu$ components are only
nonvanishing for the spatial indices, this is, we take $\eps_0\equiv 0$ for
practical purposes (recall $\eps^0 (linear~polarization)=\frac{k}{M_V}$ and 
$\eps^0=0$ for the two transverse polarizations).

\section{Spin projectors}

Next we want to find the appropriate projectors in $S=0,1,2$ in terms of the 
different combinations of the four polarization vectors. 
\begin{figure}
\begin{center}
\includegraphics{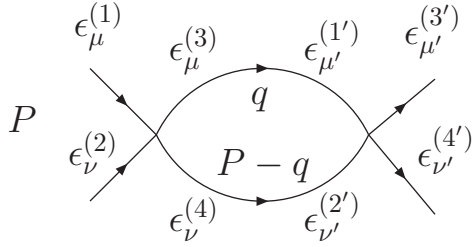}
\end{center}
\caption{Loop function for two mesons.}
\label{fig:fig3}
\end{figure}
For this purpose we look into the loop diagram of fig. \ref{fig:fig3}, where the 
interaction has been iterated to provide the second term of the Bethe Salpeter 
equation. Take the term $\eps^{(1)}_\mu \eps^{(2)}_\nu \eps^{(3)}_\mu \eps^{(4)}_\nu$ 
and iterate it. We get the structure
\begin{equation}
\eps^{(1)}_\mu \eps^{(2)}_\nu \eps^{(3)}_\mu \eps^{(4)}_\nu \eps^{(1')}_{\mu'} \eps^{(2')}_{\nu'} \eps^{(3')}_{\mu'} \eps^{(4')}_{\nu'}\ ,
\end{equation}
with the contraction of the internal indices $3,1'$ and $4,2'$ leading, after 
summing over the possible polarizations, to
\begin{equation}
(-g_{\mu\mu'}+\frac{q_\mu q_{\mu'}}{M_V^2})(-g_{\nu\nu'}+\frac{(P-q)_\nu (P-q)_{\nu'}}{M_V^2})\ ,
\end{equation}
where $P$ is the total momentum of the $\rho\rho$ system. Since $\mu,\nu,\mu',\nu'$ are all external indices, all of them are spatial and the sum above reverts into $(i,j=1,2,3)$
\begin{equation}
(\delta_{ii'}+\frac{q_i q_{i'}}{M_V^2})(\delta_{jj'}+\frac{q_j q_{j'}}{M_V^2})\ .
\label{delta}
\end{equation}
We shall work in a renormalization scheme which relies upon the function of 
$q$ apart from the two propagators in the loop function, $f(q)$, being evaluated on 
shell. The base for it can be seen in the $N/D$ method \cite{nsd,ollerulf} which 
relies upon the potential and the $T$-matrix factorized on shell in the loops 
as a result of the use of a dispersion relation on $T^{-1}$ after imposing 
unitarity. Another method of work is to recast 
$f(q)$ as $f(q_{on-shell})+(f(q)-f(q_{on-shell}))$. 
Obviously $f(q)-f(q_{on-shell})$ vanishes for $q=q_{on-shell}$, as a consequence 
of which it cancels the singularity of one meson propagator and one gets a 
diagram with a topology as in fig. \ref{fig:fig4} (the argument is the same when dealing 
with the off shell part of the other meson). This diagram gets canceled by 
tadpoles in the calculation or otherwise can be
argued to renormalize the lowest order of the $\rho\rho\to \rho\rho$ potential.

\begin{figure}
\begin{center}
\includegraphics{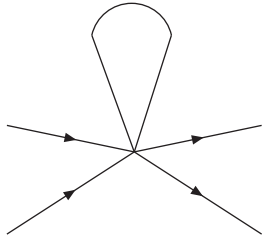}
\end{center}
\caption{Topology of the part of the diagram of fig. \ref{fig:fig3} coming from
off-shell parts of the polarization sums.}
\label{fig:fig4}
\end{figure}

The arguments above imply that we take $q_i, q_{i'}$ on shell in eq. (\ref{delta}) and $\frac{q_i q_{i'}}{M_V^2}$ is negligible and we ignore it. The argument used above is slightly different for the structure $\eps^{(1)}_\mu \eps^{(2)}_\mu \eps^{(3)}_\nu \eps^{(4)}_\nu$. Indeed, now we have the combination
\begin{equation}
\eps^{(1)}_\mu \eps^{(2)}_\mu \eps^{(3)}_\nu \eps^{(4)}_\nu \eps^{(1')}_{\mu'} \eps^{(2')}_{\mu'} \eps^{(3')}_{\nu'} \eps^{(4')}_{\nu'}\ ,
\end{equation}
and contract the indices of $3,1'$ and $4,2'$ summing over polarizations such 
that we get
\begin{equation}
(-g_{\nu\mu'}+\frac{q_\nu q_{\mu'}}{M_V^2})(-g_{\nu\mu'}+\frac{(P-q)_\nu (P-q)_{\mu'}}{M_V^2})\ .
\end{equation}
The fact that the external indices $\mu,\nu'$ are spatial does not tell us 
anything on the internal indices $\nu,\mu'$ which can also be time like. We 
can distinguish three cases:
\begin{itemize}
\item[i)] $\nu=i,\mu'=j'$ space like. On shell we get $\delta_{ij'}\delta_{ij'}$ 
with correction of $O(\vec{q}\,^2/M_V^2)$ that we neglect. 
\item[ii)] $\nu=0$, $\mu'=i$ or vice versa. We get non vanishing terms from 
$q^0(P-q)^0\vec{q}\,^2/M_V^4$ which are again of the order of $\vec{q}\,^2/M_V^2$. 
The whole term is neglected. 
\item[iii)] $\nu=0,\mu'=0$. Now the term
\begin{equation}
(-g_{00}+\frac{q_0 q_{0}}{M_V^2})(-g_{00}+\frac{(P-q)_0 (P-q)_{0}}{M_V^2})\ ,
\end{equation}
vanishes on shell up to terms of $\vec{q}\,^2/M_V^2$, which we again neglect.
\end{itemize}

The discussion has served to show that the regularization procedure is slightly different for different combinations of the polarization vectors, and hence, for the different spin terms. This implies that in dimensional regularization 
one cannot invoke exactly the same substraction constant in all channels but they cannot be too different either. We use cut off 
renormalization, and the former findings would imply possible different cut offs in different
channels but not too different, and they must be of natural size. We show results in the present paper for
different possible cut offs of natural size. There is 
another reason to allow for some freedom in the subtraction constants for different spins, since we make the approximation $\frac{q^2}{M_V^2}=0$ in the 
exchanged vectors. Given the large width of the $\rho$ meson, a consideration of the 
$\rho$ mass convolution for the four $\rho$ mesons in the amplitudes has a consequence that this quantity
would not be fully negligible for some distribution of masses in the convolution. We 
have evaluated the average value of $\frac{q^2}{M_V^2}$ for forward scattering with the  convolution of the four $\rho's$ and find $q^2/M^2_V$ of the order of $10\,\%$, 
enough to justify our calculations, but inside loops this quantity can be bigger and it is also
$s-$dependent. Once again, these effects could be accounted for by means of fine tunning of the 
substraction constants mentioned before for different channels. 

For practical purposes, in this renormalization scheme we only need the
transverse components in all cases, and in the propagators of the vector mesons
we can take

\begin{equation}
\langle T\left[ \rho_i\rho_{i'}\right] \rangle=\frac{\delta_{ii'}}{q^2-M_V^2+i\eps}
\end{equation}
or the same expression with $q\rightarrow P-q$ for the second propagator.

After this exercise it is easy to check that the three independent structures 
that upon iteration lead to the same structure are given by
\begin{eqnarray}
{\cal P}^{(0)}&=& \frac{1}{3}\eps^{(1)}_i \eps^{(2)}_i \eps^{(3)}_j \eps^{(4)}_j\nonumber\\
{\cal P}^{(1)}&=&\frac{1}{2}(\eps_i^{(1)}\eps_j^{(2)}-\eps_j^{(1)}\eps_i^{(2)})\frac{1}{2}(\eps_i^{(3)}\eps_j^{(4)}-\eps_j^{(3)}\eps_i^{(4)})\nonumber\\
{\cal P}^{(2)}&=&\lbrace\frac{1}{2}(\eps_i^{(1)}\eps_j^{(2)}+\eps_j^{(1)}\eps_i^{(2)})-\frac{1}{3}\eps_l^{(1)}\eps_l^{(2)}\delta_{ij}\rbrace\nonumber\\
&\times& \lbrace\frac{1}{2}(\eps_i^{(3)}\eps_j^{(4)}+\eps_j^{(3)}\eps_i^{(4)})-\frac{1}{3}\eps_m^{(3)}\eps_m^{(4)}\delta_{ij}\rbrace \ .
\label{eq:proji}
\end{eqnarray}
It is also easy to see that these structures project over the three different 
states of spin, $S=0,1,2$ respectively, by taking states with
 a certain third component of the spin and writing them in terms of spherical 
 vectors $\mp\frac{1}{\sqrt{2}}(\eps_1\pm i\eps_2)$ and $\eps_3$.

Although we have to keep in mind that we will be dealing with spatial 
components, it is convenient to write these projectors in covariant form such 
as to easily separate the structures that appear from the Lagrangians into the 
different spin projectors. So we write
\begin{eqnarray}
{\cal P}^{(0)}&=& \frac{1}{3}\eps_\mu \eps^\mu \eps_\nu \eps^\nu\nonumber\\
{\cal P}^{(1)}&=&\frac{1}{2}(\eps_\mu\eps_\nu\eps^\mu\eps^\nu-\eps_\mu\eps_\nu\eps^\nu\eps^\mu)\nonumber\\
{\cal P}^{(2)}&=&\lbrace\frac{1}{2}(\eps_\mu\eps_\nu\eps^\mu\eps^\nu+\eps_\mu\eps_\nu\eps^\nu\eps^\mu)-\frac{1}{3}\eps_\alpha\eps^\alpha\epsilon_\beta\epsilon^\beta\rbrace\ ,
\label{eq:projmu}
\end{eqnarray}
where the order $1,2,3,4$ in the polarization vectors is understood (this allows
us to write the expressions covariantly without complication in the
indices. We use the covariant formalism in what follows).

\section{Isospin projection}

We must now evaluate the amplitudes for the isospin states:
\begin{eqnarray}
 \vert\rho \rho, I=0\rangle&=&-\frac{1}{\sqrt{6}}\vert \rho^+(k_1\eps_1)\rho^-(k_2\eps_2)+\rho^-(k_1\eps_1)\rho^+(k_2\eps_2)+\rho^0(k_1\eps_1)\rho^0(k_2\eps_2)\rangle\nonumber\\
 \vert\rho \rho, I=1,I_3=0\rangle&=&-\frac{1}{2}\vert \rho^+(k_1\eps_1)\rho^-(k_2\eps_2)-\rho^-(k_1\eps_1)\rho^+(k_2\eps_2)\rangle\nonumber\\
 \vert\rho \rho, I=2,I_3=0\rangle&=&-\frac{1}{\sqrt{2}}\vert\frac{1}{\sqrt{6}}( \rho^+(k_1\eps_1)\rho^-(k_2\eps_2)+\rho^-(k_1\eps_1)\rho^+(k_2\eps_2))\nonumber\\&\hspace{0.3cm}&-\sqrt{\frac{2}{3}}\rho^0(k_1\eps_1)\rho^0(k_2\eps_2)\rangle
 \label{states}
\end{eqnarray}
Note that we are using the unitary normalization \cite{npa}, with an extra 
factor $\frac{1}{\sqrt{2}}$ such that when summing over intermediate states of 
identical particles we obtain the resolution of the identity, i.e. 
$\frac{1}{2}\sum_{q}\vert\rho^0(\vec{q})\rho^0(-\vec{q})\rangle\langle\rho^0(\vec{q})\rho^0(-\vec{q})\vert=1$. 
One must correct the final amplitudes for the "wrong" normalization on the 
external legs, with a global normalization factor that does not affect the search 
for the poles or the energy dependence of the amplitudes. We also take the phase
convention $\vert\rho^+\rangle=-\vert 1,1\rangle$ of isospin.

By using the isospin wave functions and the Lagrangian of eq. (\ref{lcont}) we obtain for $I=1$
\begin{equation}
t^{(I=1)}=3g^2(\eps_\mu\eps_\nu\eps^\mu\eps^\nu-\eps_\mu\eps_\nu\eps^\nu\eps^\mu)\ ,
\end{equation}
which according to the spin projection operators of eq. (\ref{eq:projmu}) only 
has the $S=1$ component, consistent with the rule $L+S+I=even$. Thus we have
\begin{equation}
t^{(I=1,S=1)}\equiv 6g^2\ .
\label{t11}
\end{equation}
The interaction is repulsive, however we still have to evaluate the contribution
 from the vector exchange mechanisms. In $I=0$ we get the amplitude
\begin{equation}
t^{(I=0)}=2g^2\lbrace
2\eps_\mu\eps^\mu\eps_\nu\eps^\nu-\eps_\mu\eps_\nu\eps^\mu\eps^\nu-\eps_\mu\eps_\nu\eps^\nu\eps^\mu\rbrace\ ,
\label{iscero}
\end{equation}
which by means of the spin projection structures leads to 
\begin{eqnarray}
t^{(I=0,S=0)}&=&8g^2\nonumber\\
t^{(I=0,S=2)}&=&-4g^2\ .
\label{t02}
\end{eqnarray}
We can see that the interaction in the $I=0,S=0$ channel is repulsive but the 
one in $S=2$ is attractive. We still need, however, the contribution of the 
vector exchange terms. Note again that according to the rule $L+S+I=even$ we 
do not get contribution of $S=1$ for $I=0$. In $I=2$ we obtain the amplitude
\begin{equation}
t^{(I=2)}=-g^2(2\eps_\mu\eps^\mu\eps_\nu\eps^\nu
-\eps_\mu\eps_\nu\eps^\mu\eps^\nu -\eps_\mu\eps_\nu \eps^\nu\eps^\mu)\ ,
\label{t2}
\end{equation}
which projected over spin states leads to
\begin{eqnarray}
t^{(I=2,S=0)}&=&-4g^2\nonumber\\
t^{(I=2,S=2)}&=&2g^2\ .
\label{t22}
\end{eqnarray}
\section{Vector exchange terms}

\begin{figure}
\begin{center}
\includegraphics{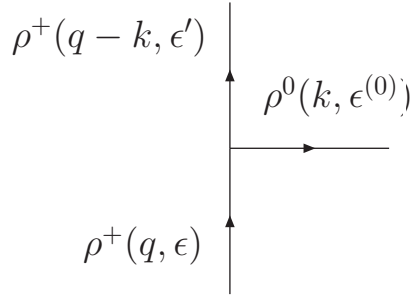}
\end{center}
\caption{Three vector vertex diagram.}
\label{fig:fig5}
\end{figure}

\begin{figure}
\begin{center}
\includegraphics{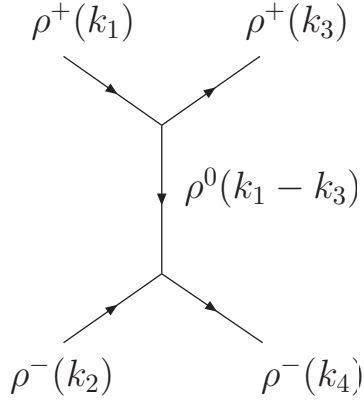}
\end{center}
\caption{Vector exchange diagram for $\rho^+\rho^-\to\rho^+\rho^-$.}
\label{fig:fig6}
\end{figure}

From the Lagrangian of eq.(\ref{l3V}) we get the three vector vertex depicted 
in fig. \ref{fig:fig5}.
The vertex function corresponding to the diagram of fig. \ref{fig:fig3} is given 
by
\begin{eqnarray}
-it^{(3)}=&-&\sqrt{2}g\lbrace (ik_\mu \eps^{(0)}_\nu -ik_\nu \eps^{(0)}_\mu)\eps^\mu\eps'^{\nu}\nonumber\\&+&(-iq_\mu\eps_\nu+iq_\nu\eps_\mu)\eps'^{\mu}\eps^{(0)\nu}\nonumber\\&+&(i(q-k)_\mu \eps'_\nu-i(q-k)_\nu \eps'_\mu)\eps^{(0)\mu}\eps^\nu\rbrace\ ,
\label{vertexfig3}
\end{eqnarray}
with this basic structure we can readily evaluate the amplitude of the diagram 
of fig. \ref{fig:fig6} and we obtain 

\begin{eqnarray}
-it^{(ex)}=&-&\sqrt{2} g\lbrace (i(k_1-k_3)_\mu \eps^{(0)}_\nu -i(k_1-k_3)_\nu\eps^{(0)}_\mu)\eps^{(1)\mu}\eps^{(3)\nu}\nonumber\\&+&(-ik_{1\mu}\eps^{(1)}_\nu +ik_{1\nu}\eps^{(1)}_\mu)\eps^{(3)\mu}\eps^{(0)\nu}+(ik_{3\mu}\eps^{(3)}_\nu -ik_{3\nu}\eps^{(3)}_\mu)\eps^{(0)\mu}\eps^{(1)\nu}\rbrace\nonumber\\ &\times&\frac{i}{(k_1-k_3)^2-M_\rho^2+i\eps}\nonumber\\
&\times&(-\sqrt{2})g\lbrace (-i(k_2-k_4)_{\mu'} \eps^{(0)}_{\nu'} +i(k_2-k_4)_{\nu'}\eps^{(0)}_{\mu'})\eps^{(4)\mu'}\eps^{(2)\nu'}\nonumber\\&+&(ik_{4\mu'}\eps^{(4)}_{\nu'} -ik_{4\nu'}\eps^{(4)}_{\mu'})\eps^{(2)\mu'}\eps^{(0)\nu'}+(ik_{2\mu'}\eps^{(2)}_{\nu'} +ik_{2\nu'}\eps^{(2)}_{\mu'})\eps^{(0)\mu'}\eps^{(4)\nu'}\rbrace\ .
\label{exchange}
\end{eqnarray}

At this point we must recall that the three momenta of the external particles 
is small and neglected, so that we keep only spatial components of the 
polarization vectors. As a consequence, the term $(k_1-k_3)^2$ in the $\rho$ 
propagator is neglected. Similarly all terms of the type $k_{i\mu}\eps^\mu$, 
$k_{i\mu}\eps^{'\mu}$ can be neglected and only the terms 
$k_{i\mu}\eps^{(0)\mu}$ remain, since the exchanged vector can be time like 
(the only component that survives). As a consequence our amplitude gets much 
simplified and we obtain
\begin{equation}
-it^{(ex)}=2i\frac{g^2}{M_\rho^2}(k_1\cdot k_4 +k_3\cdot k_4 +k_1\cdot k_2+k_2\cdot k_3)\eps_\mu\eps_\nu\eps^\mu\eps^\nu\ ,
\label{exch}
\end{equation}
with a unique spin structure, which can be recast, using momentum conservation 
into
\begin{equation}
t^{(ex)}=-\frac{4g^2}{M_\rho^2}(\frac{3}{4}s-M_\rho^2)\eps_\mu\eps_\nu\eps^\mu\eps^\nu\ .
\label{exchs}
\end{equation}
The approximations done here are the same ones that one would do for the 
interaction of a vector with a pseudoscalar and which lead to the local chiral 
Lagrangian of \cite{birse,lutzaxial,luisaxial} used to generate axial vector 
mesons in \cite{lutzaxial,luisaxial}. This example serves to place our 
approximations in a due perspective, since such approximations are implicit in 
most of the effective chiral Lagrangians used in the literature \cite{ulf}, 
which can be deduced from the formalism of the hidden gauge Lagrangians that 
we are using here. 

\begin{figure}
\begin{center}
\includegraphics{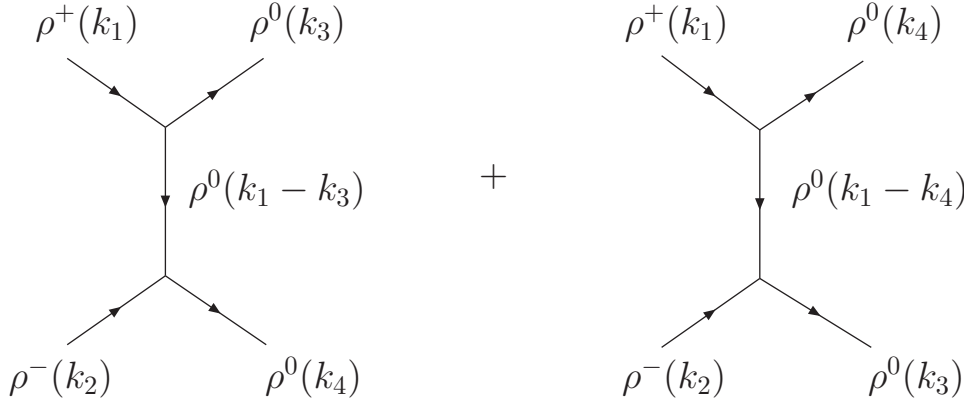}
\end{center}
\caption{$t$ and $u$ channel exchange of vector.}
\label{fig:fig7}
\end{figure}

Before proceeding further one must evaluate the amplitudes for the different 
isospin states and we obtain
\begin{equation}
t^{(ex,I=1)}=-\frac{2g^2}{M_\rho^2}(\frac{3}{4}s-M_\rho^2)(\eps_\mu\eps_\nu\eps^\mu\eps^\nu-\eps_\mu\eps_\nu\eps^\nu\eps^\mu)\ ,
\label{exchange1}
\end{equation}
which already projects over $S=1$, as it should, such that 
\begin{equation}
t^{(ex,I=1,S=1)}=-4g^2(\frac{3s}{4M_\rho^2}-1)\ .
\label{exchange11}
\end{equation}
The case of $I=0,2$ is more subtle because, unlike the case of the contact term, 
we have now a $t$ and $u$ exchange channel, see fig. \ref{fig:fig7}.
When the two diagrams are considered we obtain
\begin{equation}
t^{(ex,I=0)}=-\frac{4g^2}{M_\rho^2}(\frac{3}{4}s-M_\rho^2)(\eps_\mu\eps_\nu\eps^\mu\eps^\nu+\eps_\mu\eps_\nu\eps^\nu\eps^\mu)\ ,
\label{exchange0}
\end{equation}
which upon spin projection leads to
\begin{eqnarray}
t^{(ex,I=0,S=0)}&=&-8g^2(\frac{3s}{4M_\rho^2}-1)\nonumber\\
t^{(ex,I=0,S=2)}&=&-8g^2(\frac{3s}{4M_\rho^2}-1)\ .
\label{exch0002}
\end{eqnarray}
Similarly for $I=2$ we obtain
\begin{equation}
t^{(ex,I=2)}=\frac{2g^2}{M_\rho^2}(\frac{3}{4}s-M_\rho^2)(\eps_\mu\eps_\nu\eps^\mu\eps^\nu+\eps_\mu\eps_\nu\eps^\nu\eps^\mu)\ ,
\label{exchange2}
\end{equation}
which upon spin projection leads to
\begin{eqnarray}
t^{(ex,I=2,S=0)}&=&4g^2(\frac{3s}{4M_\rho^2}-1)\nonumber\\
t^{(ex,I=2,S=2)}&=&4g^2(\frac{3s}{4M_\rho^2}-1)\ .
\label{exch2022}
\end{eqnarray}
The results obtained with the contact term and the $\rho$-exchange mechanism 
provide the kernel, or potential $V$, to be used in the Bethe Salpeter equation 
in its on-shell factorized form,
\begin{equation}
T= \frac{V}{1-VG}\ ,
\label{Bethe}
\end{equation}
for each spin-isospin channel independently, where $G$ is the two $\rho$ loop 
function in the approximation of neglecting the on-shell three momenta
\begin{equation}
G=i\int \frac{d^4 q}{(2\pi)^4}\frac{1}{q^2-m_\rho^2+i\eps}\frac{1}{(P-q)^2-m_\rho^2+i\eps}\ ,
\label{loop}
\end{equation}
which upon using dimensional regularization can be recast as
\begin{eqnarray}
G&=&{1 \over 16\pi ^2}\biggr( \alpha +Log{m_1^2 \over \mu ^2}+{m_2^2-m_1^2+s\over 2s}
  Log{m_2^2 \over m_1^2}\nonumber\\ 
  &+&{p\over \sqrt{s}}\Big( Log{s-m_2^2+m_1^2+2p\sqrt{s} \over -s+m_2^2-m_1^2+
  2p\sqrt{s}}+Log{s+m_2^2-m_1^2+2p\sqrt{s} \over -s-m_2^2+m_1^2+  2p\sqrt{s}}\Big)\biggr)\ ,
  \label{dimreg}
\end{eqnarray}
 where P is the total four-momentum of the two mesons, $p$ is the three-momentum 
 of the mesons in the center of mass frame and  $m_1=m_2=m_\rho$, or using a cut 
 off as
 \begin{equation}
G=\int_0^{q_{max}} \frac{q^2 dq}{(2\pi)^2} \frac{\omega_1+\omega_2}{\omega_1\omega_2 [{(P^0)}^2-(\omega_1+\omega_2)^2+i\epsilon]   } \ ,\label{loopcut}
\end{equation}
 where $q_{max}$ stands for the cut off, $\omega_i=(\vec{q}\,^2_i+m_i^2)^{1/2}$ 
 and the center-of-mass energy ${(P^0)}^2=s$. 
 The potential $V$ obtained summing the lowest order $T$ matrices obtained from 
 the contact term and $\rho$ exchange are summarized in Table \ref{tab:V}.\\
 \begin{table}[h]
 \begin{center}
\begin{tabular}{c|c|c|c|c}
$I$&$S$&Contact&Exchange&$\sim Total[I^G(J^{PG})]$\\
\hline
\hline
$1$&$1$&$6g^2$&$-4g^2(\frac{3s}{4M_\rho^2}-1)$&$-2g^2[1^+(1^{+-})]$\\
\hline
$0$&$0$&$8g^2$&$-8g^2(\frac{3s}{4M_\rho^2}-1)$&$-8g^2[0^+(0^{++})]$\\
\hline
$0$&$2$&$-4g^2$&$-8g^2(\frac{3s}{4M_\rho^2}-1)$&$-20g^2[0^+(2^{++})]$\\
\hline
$2$&$0$&$-4g^2$&$4g^2(\frac{3s}{4M_\rho^2}-1)$&$4g^2[0^+(2^{++})]$\\
\hline
$2$&$2$&$2g^2$&$4g^2(\frac{3s}{4M_\rho^2}-1)$&$10g^2[0^+(2^{++})]$\\
\hline
\end{tabular} 
\end{center}
\caption{$V$ for the different spin isospin channels}
\label{tab:V}
\end{table}
In Table \ref{tab:V} we have written in the last column the quantum numbers of 
the state and the approximate strength of the potential calculated at the 
$\rho\rho$ threshold to get an idea of the weight of the interaction. We observe
 attraction in the $I,S=1,1;0,0$ and $0,2$ channels and repulsion in $2,0;2,2$. 
 We, thus, can not generate $I=2$ low lying states from this $\rho\rho$ interaction. We find a weak attraction for the $I,S=1,1$ with $1^+(1^{+-})$ quantum numbers
  and then a strong attraction for $I,S=0,0$ and a much larger attraction for 
  $I,S=0,2$, anticipating that if the interaction leads to a bound $\rho\rho$ 
  state with $I,S=0,0$ it will necessarily lead to a much deeper bound 
  $I,S=0,2$ state, a trend actually followed by the $f_0(1370)$ and $f_2(1270)$ 
  resonances. The case of $I,S=1,1$ with $1^+(1^{+-})$ quantum numbers is 
  special. These are the quantum numbers of the $b_1(1235)$. This state is 
  generated dynamically from the interaction of vector mesons with 
  pseudoscalars, the $KK^*$ channel being the most important one 
  \cite{luisaxial}. The weak interaction of the possible $\rho\rho$ component of 
  this state and the fact the $\rho\rho$ threshold is $300\,MeV$ above the mass 
  of the $b_1(1235)$ anticipate that the $\rho\rho$ channel investigated here 
  will have little effect modifying the results obtained for that resonance 
  from the dynamics of the $KK^*$ interaction. The weak attraction in this channel 
  does not support a $\rho\rho$ bound state but could lead to a broad resonance at 
  higher energies that we do not investigate here.

One may wonder about mixing the $\omega$ channel with the $\rho$. One can
easily see that there are no contact terms with $\omega$ and the three vector
vertices mixing $\omega$ with $\rho$ are also forbidden since
$\rho\omega\omega$ violates isospin and $\rho\rho\omega$ violates $G$ parity. 

\begin{figure}
\begin{center}
\includegraphics{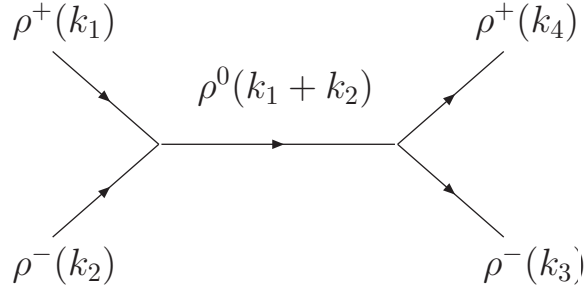}
\end{center}
\caption{$S$-channel $\rho$ exchange diagram.}
\label{fig:fig8}
\end{figure}

The formalism that we are using is also allowed for $s$-channel $\rho$ exchange 
and we can have the diagram of fig. \ref{fig:fig8}. 
By performing similar approximations as done before we obtain an amplitude 
only in $I=1,S=0$ of the type
\begin{equation}
t^{(s)}=24g^2\frac{1}{(k_1+k_2)^2-M_\rho^2}\vec{k}_1\cdot\vec{k}_3\ ,
\label{schannel}
\end{equation}
which is a p-wave amplitude and repulsive. Note it also satisfies $L+S+I=even$.
\section{Convolution due to the $\rho$ mass distribution}

The strong attraction in the $I,S=0,0;0,2$ channels will produce $\rho\rho$ 
bound states and thus with no width within the model. However, this is not 
strictly true because the $\rho$ has a large width or equivalently a mass 
distribution that allows the states obtained to decay in $\rho\rho$ for the 
low mass components of the $\rho$ mass distribution. To take this into account 
we follow the traditional method which is to convolute the $G$ function for the 
mass distributions of the two $\rho$ mesons \cite{luishideko} replacing the $G$ 
function by $\tilde{G}$ as follows
\begin{eqnarray}
\tilde{G}(s)&=& \frac{1}{N^2}\int^{(m_\rho+2\Gamma_\rho)^2}_{(m_\rho-2\Gamma_\rho)^2}d\tilde{m}^2_1(-\frac{1}{\pi}) {\cal I}m\frac{1}{\tilde{m}^2_1-m^2_\rho+i\Gamma\tilde{m}_1}\nonumber\\
&\times&\int^{(m_\rho+2\Gamma_\rho)^2}_{(m_\rho-2\Gamma_\rho)^2}d\tilde{m}^2_2(-\frac{1}{\pi}) {\cal I}m\frac{1}{\tilde{m}^2_2-m^2_\rho+i\Gamma\tilde{m}_2} G(s,\tilde{m}^2_1,\tilde{m}^2_2)\ ,
\label{Gconvolution}
\end{eqnarray}
with
\begin{equation}
N=\int^{(m_\rho+2\Gamma_\rho)^2}_{(m_\rho-2\Gamma_\rho)^2}d\tilde{m}^2_1(-\frac{1}{\pi}){\cal I}m\frac{1}{\tilde{m}^2_1-m^2_\rho+i\Gamma\tilde{m}_1}\ ,
\label{Norm}
\end{equation}
where $\Gamma_\rho=146.2\, MeV$ and for $\Gamma\equiv\Gamma(\tilde{m})$ we take the $\rho$ width for the decay into the pions in $p$-wave
\begin{equation}
\Gamma(\tilde{m})=\Gamma_\rho (\frac{\tilde{m}^2-4m^2_\pi}{m^2_\rho-4m^2_\pi})^{3/2}\theta(\tilde{m}-2m_\pi)
\label{gamma}
\end{equation}
The use of $\tilde{G}$ in eq. (\ref{Bethe}) provides a width to the states 
obtained. 
\section{Results}

In the first step we calculate the $T$ matrix for the scattering of $\rho\rho$ 
in the two channels $I,S=0,0;0,2$ which experience the largest attraction 
according to Table \ref{tab:V}. We consider the potential coming from the 
contact and exchange term, not the approximate sum shown in the table. For 
reasonable choices of the cut off, $q_{max}$, of the order of $1\,GeV$ we 
always find bound states for both sets of quantum numbers, easily visible 
since $T$ goes to infinity at values of $\sqrt{s}$ smaller than two $\rho$-meson 
masses. In Table \ref{tab:T2} we show the energies of the bound states for 
different values of the cut off when we take a fixed $\rho$ mass equal to its 
nominal mass. 

\begin{table}[h]
\begin{center}
\begin{tabular}{c|c|c|c}
$I$&$S$&$\sqrt{s}~(MeV)[q_{max}=875\,MeV/c]$&$\sqrt{s}~(MeV)[q_{max}=1000\, MeV/c]$\\
\hline
\hline
$0$&$0$&$1512$&$1491$\\
\hline
$0$&$2$&$1255$&$1195$\\
\hline
\end{tabular}
\end{center}
\caption{Pole positions for the two different channels}
\label{tab:T2}
\end{table}
We have used two values of the cut off around $1\,GeV/c$, $875\,MeV/c$ and $1000\,MeV/c$. 
What we see is that in both cases, and for higher values of $q_{max}$, one gets 
bound states for both $S=0,S=2$, and the binding of the $S=2$ state is bigger
 than for $S=0$. Since the strength of the potential for $S=2$ is much bigger 
 than for $S=0$, we also see that the binding of the tensor state is more 
 sensitive to the cut off than that of the scalar state. Yet, reasonable 
 changes of $q_{max}$ around $1\,GeV$ revert into changes of about $50\,MeV$ in 
 the binding for the tensor state and about $20\,MeV$ for the scalar state. As 
 usually done in this kind of calculations, once one shows the qualitative 
 features of the states obtained, one can do some fine tuning of the 
 parameters, only $q_{max}$ in the present case, in order to match the energy of a 
 certain state. In this case we choose the $f_2(1270)$ tensor state, since its 
 mass is very precisely determined from different experiments \cite{pdg} at 
 $1275\,MeV$. 
 Unlike the case of the $f_2(1270)$ state which has a well defined mass, the 
 $f_0(1370)$ has a large dispersion of values in the $PDG$ \cite{pdg} to the point 
 that they quote a mass $1200-1500\, MeV$ as their average.

As to the width, in our calculation it is obviously zero in both cases since we
have obtained $\rho\rho$ bound states. Experimentally we have
$\Gamma(f_2(1270))=184.4^{+ 3.9}_{- 2.5}\,MeV$ and
$\Gamma(f_0(1370))=200-500\,MeV$ according to \cite{pdg}. Let us see if we can
find a reasonable width for these states once we take into account the $\rho$
mass distribution as described in the former section. 

\begin{figure}
\includegraphics[width=16.2cm]{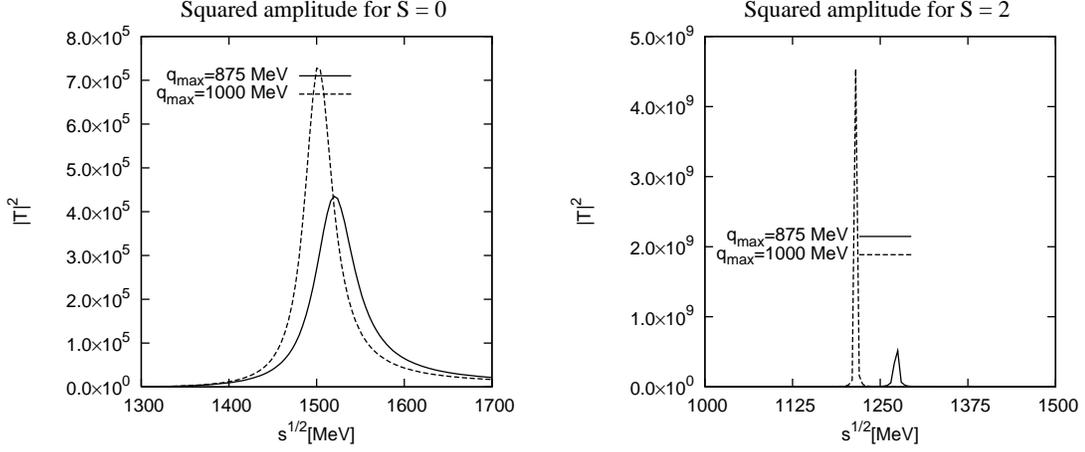}
\caption{$\vert T\vert^2$ taking into account the $\rho$ mass distribution for
$S=0$ and $S=2$.}
\label{fig:convo}
\end{figure}

In fig. \ref{fig:convo} we show the results for $\vert T\vert^2$ obtained by
considering the $\rho$ mass distribution. We show the results for the two cut
offs of Table \ref{tab:T2}. As we can see in the figure, the matching of the mass of the $f_2(1270)$ is obtained with a 
 cut off $q_{max}=875\,MeV/c$. Then we obtain $1532\,MeV$ for the energy of the 
 $S=0$ state that we would like to associate to the $f_0(1370)$.  Given,  
 large dispersion of masses of the $f_0(1370)$, the results obtained by us would be consistent with the 
 present experimental observation. 

We see that $\vert T\vert^2$  has a good Breit Wigner distribution in both 
cases, with a peak around the masses shown in Table \ref{tab:T2}, but changed 
slightly. However, the 
widths are relatively small. For the tensor state one finds a width of about 
$2-3\,MeV$ and for the scalar state the width is about $50-75\, MeV$, depending 
on the cut off. 

We see that the consideration of the $\rho$ mass distribution leads indeed to a
width of the states, but it is still very small compared with experiment,
particularly for the tensor state. Clearly, there must be other sources of
imaginary part. The likely candidate for the decay must be two pions, and
indeed this accounts for $84.7\,$\% of the total width in the case of the
$f_2(1270)$. The case of the $f_0(1370)$ is less clear since the two pion
fraction can be of the order of $20\,$\% \cite{pdg} while the $4\pi$ fraction
could be dominant. 

In the next section we address this problem and study the mechanisms that lead 
to the two pions decay of the two $\rho$ system.

\section{Consideration of the two pion decay mode}

The results obtained are interesting in as much as we are obtaining the two 
states $f_0(1370)$ and $f_2(1270)$ qualitatively, with the important fact that 
the $f_2(1270)$ state is more bound than the $f_0(1370)$. In this section we
take into account the diagrams that couple $\rho\rho$ to $\pi\pi$, thus mixing
the $\pi\pi$ channel with the $\rho\rho$ and allowing the states obtained to
develop a width from decay into $\pi\pi$.

\begin{figure}
\begin{center}
\includegraphics[width=16.2cm]{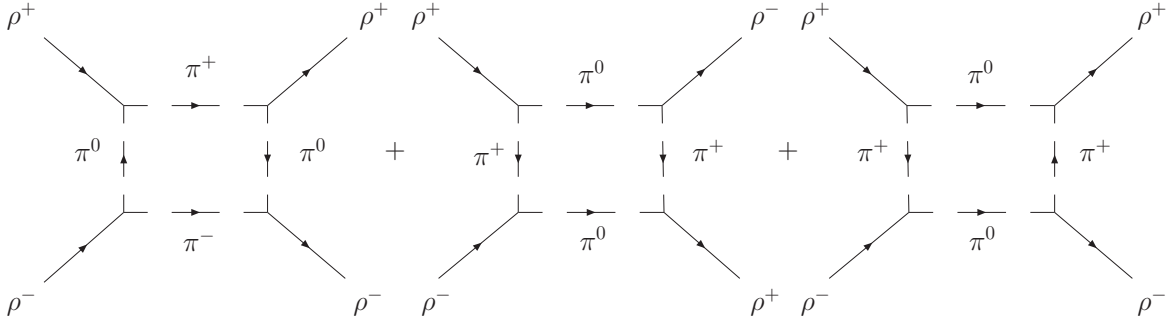}
\end{center}
\caption{Diagrams considered for $ \rho\rho\to\pi\pi$.}
\label{fig:fig9}
\end{figure}

The $\pi\pi$ interaction at these energies away from $\pi\pi$ threshold, in 
$L=0$ and $L=2$ that we will need for states of $J=0,2$ respectively, is rather 
weak compared with the one of the $\rho\rho$ interaction. Furthermore,
the energies of the two resonances under discussion are close to the two $\rho$
meson threshold and far away for that of $\pi \pi$. Hence, this latter channel
cannot have much weight in the wave function of these resonances. It is, thus,
 unnecessary to treat 
the $\pi\pi$ as a coupled channel and one can simplify the work by computing the 
diagrams for $\rho\rho\to\rho\rho$ mediated by two pion exchange depicted in 
fig. \ref{fig:fig9} for $\rho^+\rho^-\to\rho^+\rho^-$. 
If we introduce these new terms as part of the $\rho\rho$ interaction and 
iterate them through the Bethe Salpeter equation of eq. (\ref{Bethe}), we 
generate all terms with transition of $\rho\rho$ to $\pi\pi$ and neglect terms 
containing the $\pi\pi\to\pi\pi$ interaction that we considered weaker. 

\begin{figure}
\begin{center}
\includegraphics[width=7cm]{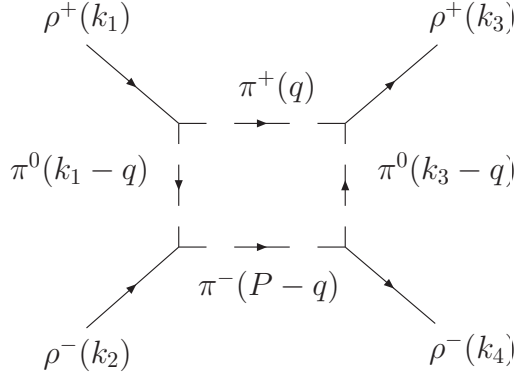}
\end{center}
\caption{ Detail of one of the diagrams of fig. \ref{fig:fig9}.}
\label{fig:fig10}
\end{figure}
The evaluation of the box diagram in fig. \ref{fig:fig10}, where the momenta are 
explicitly shown, is straightforward. One needs the $\rho\pi\pi$ couplings which 
are provided within the same framework of the hidden gauge formalism
\cite{hidden1,hidden2} by means 
of the Lagrangian
\begin{equation}
{\cal L}_{V\Phi\Phi}=-ig\langle V^\mu[\Phi,\partial_\mu \Phi]\rangle\ .
\label{lVPP}
\end{equation}
We have:
\begin{eqnarray}
-it^{(\pi\pi)}&=&\int\frac{d^4 q}{(2\pi)^4}(-i)(\sqrt{2}g)^4(q-k_1+q)^\mu\eps^{(1)}_\mu \nonumber\\
&\times& i(k_1-q+P-q)_\nu\eps^{(2)\nu} i(k_3-q-q)_\alpha\eps^{(3)\alpha}
(-i)(q-k_3-P+q)_\beta\eps^{(4)\beta}\nonumber\\&\times&\frac{i}{q^2-m^2_\pi+i\eps}\,\frac{i}{(k_1-q)^2-m^2_\pi+i\eps}\nonumber\\
&\times&\,\frac{i}{(P-q)^2-m^2_\pi+i\eps}\,\frac{i}{(k_3-q)^2-m^2_\pi+i\eps}\ .
\label{tbox}
\end{eqnarray}
By making again the approximation that all the polarization vectors are spatial, 
we can rewrite the amplitude as
\begin{eqnarray}
-it^{(\pi\pi)}&=&(\sqrt{2}g)^4\int\frac{d^4q}{(2\pi)^4}16\,q_i q_jq_lq_m \, \eps^{(1)}_i\eps^{(2)}_j\eps^{(3)}_l\eps^{(4)}_m\nonumber\\&\times&\frac{1}{q^2-m^2_\pi+i\eps}\,\frac{1}{(k_1-q)^2-m^2_\pi+i\eps}\nonumber\\
&\times&\frac{1}{(P-q)^2-m^2_\pi+i\eps}\,\frac{1}{(k_3-q)^2-m^2_\pi+i\eps}\ .
\label{tbox1}
\end{eqnarray}
Since the integral is logarithmically divergent, and in the absence of data 
to fit the subtraction constant if using dimensional regularization, we 
regularize it with a cut off in the three momentum, which should be of the 
order of $1\,GeV$, the basic scale at the energies that we are working. This 
requires to perform the $q^0$ integration analytically, which is easily done by 
means of the residues, caring to divide exactly by factors with undefined 
polarity ($\pm i\eps$ in factors of the denominator). The procedure proves more 
practical in this case than using the Feynman parametrization, which requires 
three integrals, while here we need only one. Furthermore, the cuts, or sources 
of imaginary part show up explicitly and allow one to keep control in the 
numerical evaluation. After some algebraic manipulation we obtain
\begin{eqnarray}
V^{(\pi\pi)}&=&(\sqrt{2}g)^4\, \left( \eps^{(1)}_i\eps^{(2)}_i\eps^{(3)}_j\eps^{(4)}_j+\eps^{(1)}_i\eps^{(2)}_j\eps^{(3)}_i\eps^{(4)}_j+\eps^{(1)}_i\eps^{(2)}_j\eps^{(3)}_j\eps^{(4)}_i\right)\nonumber\\
&\times& \frac{8}{15\pi^2}\int^{q_{max}}_0\,dq\, \vec{q}\,^6\,\lbrace10\omega^2-(k_3^0)^2\rbrace \frac{1}{\omega^3}\left( \frac{1}{k_1^0+2\omega}\right) ^2\frac{1}{P^0+2\omega}\nonumber\\
&\times&\frac{1}{k^0_1+\frac{\Gamma}{4}-2\omega+i\eps}\,
\frac{1}{k^0_1-\frac{\Gamma}{4}-2\omega+i\eps}\,\frac{1}{P^0-2\omega+i\eps}\ ,
\label{Vbox}
\end{eqnarray}
with $\omega=\sqrt{\vec{q}\,^2+ m_\pi ^2}$. We see the two sources of imaginary part in the cuts 
$k_1^0\pm\frac{\Gamma}{4}-2\omega=0$ and $P^0-2\omega=0$, corresponding to 
$\rho\to\pi\pi$ and $\rho\rho\to\pi\pi$. The diagram of Fig. \ref{fig:fig10} leads to a 
double pole for $\rho\to\pi\pi$, a decay channel that is open at the energies that we study . But when the mass distribution
of the $\rho$ is considered in the evaluation of the amplitude, the degeneracy of the pole is removed. We have performed such
a calculation, but have observed that a simpler approach, and accurate enough for our purposes, is to substitute the double pole $(1/(k_1^0-2\omega+i\eps))^2$, which appears in the calculation with fixed $\rho$ masses, by $(1/(k_1^0-2\omega +\frac{\Gamma}{4}+i\eps))\,(1/(k_1^0-2\omega -\frac{\Gamma}{4}+i\eps))$ to approximatly account for the dispersion of $\rho$ masses in the convolution. 
We see in practice that the results barely change if we put there $\Gamma/2$ 
instead of $\Gamma/4$ or some other reasonable number of the size of the $\rho$ 
width. 

The spin structure projects over $S=0,S=2$, not over $S=1$, which is obvious
since the parity of the $\rho\rho$ system for $\rho$ in $s$-wave is positive
which forces the two pions in $L=0,2$, equivalent to total $J$ since the pions
have no spin. Hence we find only the $0^+,2^+$ quantum numbers. 

The next step is to evaluate the diagrams shown in Fig. \ref{fig:fig9} for  $\rho^+\rho^-\to\rho^+\rho^-$ in the case of  $\rho^+\rho^-\to\rho^0\rho^0$, 
etc, to obtain the isospin combinations. We are only interested in $I=0$, for 
the two states found in the former sections. So we obtain finally
\begin{eqnarray}
t^{(2\pi,I=0,S=0)}&=&20\,\tilde{V}^{(\pi\pi)}\nonumber\\
t^{(2\pi,I=0,S=2)}&=&8\,\tilde{V}^{(\pi\pi)}
\end{eqnarray}
where $\tilde{V}^{(\pi\pi)}$ is given by the eq. (\ref{Vbox}) after removing the
polarization vectors.

The integral of eq. (\ref{Vbox}) is logarithmically divergent, a divergence 
that can be smoothly regularized with a cut off as we have done before. We 
have checked that with the former cut off the real part obtained from 
$V^{(\pi\pi)}$ is fairly smaller than that obtained from the $\rho\rho$ 
potentials of Table \ref{tab:V}. On the other hand, there is also another 
source of real part from the box diagram involving the $\rho\omega\pi$ 
anomalous coupling, which has a similar structure ($\vec{q}\,^6$ factor 
in the integrand), and is also smaller than the potentials of Table \ref{tab:V}
and of opposite sign to $V^{(\pi\pi)}$. Altogether, we neglect the real parts 
of these box diagrams and take the real part of the potential from the tree 
level potential of Table \ref{tab:V}. In the next section we come back to these issues with a 
detailed evaluation. However, $V^{(\pi\pi)}$ leads now to 
a large imaginary part 
of the resonances because of the large phase space for $\pi\pi$ decay. The 
largest piece of the imaginary part comes from the factor $(P^0-2\omega+i\eps)^{-1}$. 
Since we are concerned about the width of the resonances, we, thus, consider the 
$\pi$ exchange between two $\rho$ mesons in the $t$ channel as mostly off shell 
and implement empirical form factors used in the decay of vector mesons in 
\cite{titov1},\cite{titov2}. We use 
\begin{equation}
F(q)=\frac{\Lambda^2-m^2_\pi}{\Lambda^2-(k-q)^2}
\label{formfactor}
\end{equation}
in each $\rho\to\pi\pi$ vertex with 
\begin{eqnarray}
k^0 = \frac{\sqrt{s}}{2}\hspace{0.5cm} \vec{k}=0\hspace{0.5cm} q^0=\frac{\sqrt{s}}{2}\ ,
\end{eqnarray}
and $\vec{q}$ the running variable in the integral. 

We shall evaluate the results for different values of $\Lambda$ around
$1200-1300\,MeV$, which are the values chosen in \cite{titov1},\cite{titov2}. 
We also implement a global cut off of $q_{max}=1.2\,GeV$ in the integral of 
eq. (\ref{Vbox}), although the form factors already provide fast convergence 
around that region.

\section{Consideration of the crossed-$\pi\pi$ box diagrams and the two omega intermediate state}

\begin{figure}
\begin{center}
\includegraphics[width=7cm]{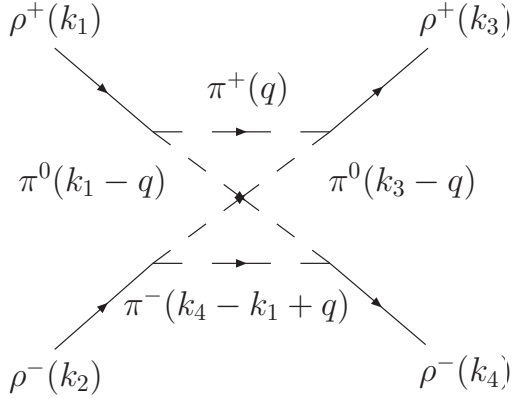}
\end{center}
\caption{ Crossed-box diagram for the four pion decay mode.}
\label{fig:boxcross}
\end{figure}

We can also have the crossed diagram of Fig. \ref{fig:boxcross}. 
By following identical steps as for the diagram of Fig. \ref{fig:fig10} we obtain at 
the end the expression
\begin{eqnarray}
\tilde{V}^{(c,\,\pi\pi)}&=&  \frac{16\,g^4}{15\pi^2}\int^{q_{max}}_0\,dq\, \vec{q}\,^6\,\lbrace20\omega^2-(k_1^0)^2\rbrace \frac{1}{\omega^3}\left( \frac{1}{k_1^0+2\omega}\right) ^3\nonumber\\
&\times&\frac{1}{k^0_1+\frac{\Gamma}{4}-2\omega+i\eps}\,
\frac{1}{k^0_1-\frac{\Gamma}{4}-2\omega+i\eps}\,\frac{1}{k_1^0-2\omega+i\eps}
\label{Vboxcross}
\end{eqnarray}
and 
\begin{eqnarray}
t^{(2\pi(c),I=0,S=0)}&=&5\,\tilde{V}^{(c,\,\pi\pi)}\nonumber\\
t^{(2\pi(c),I=0,S=2)}&=&2\,\tilde{V}^{(c,\,\pi\pi)}\ .
\end{eqnarray}
 
\begin{figure}
\begin{center}
\includegraphics[width=13cm]{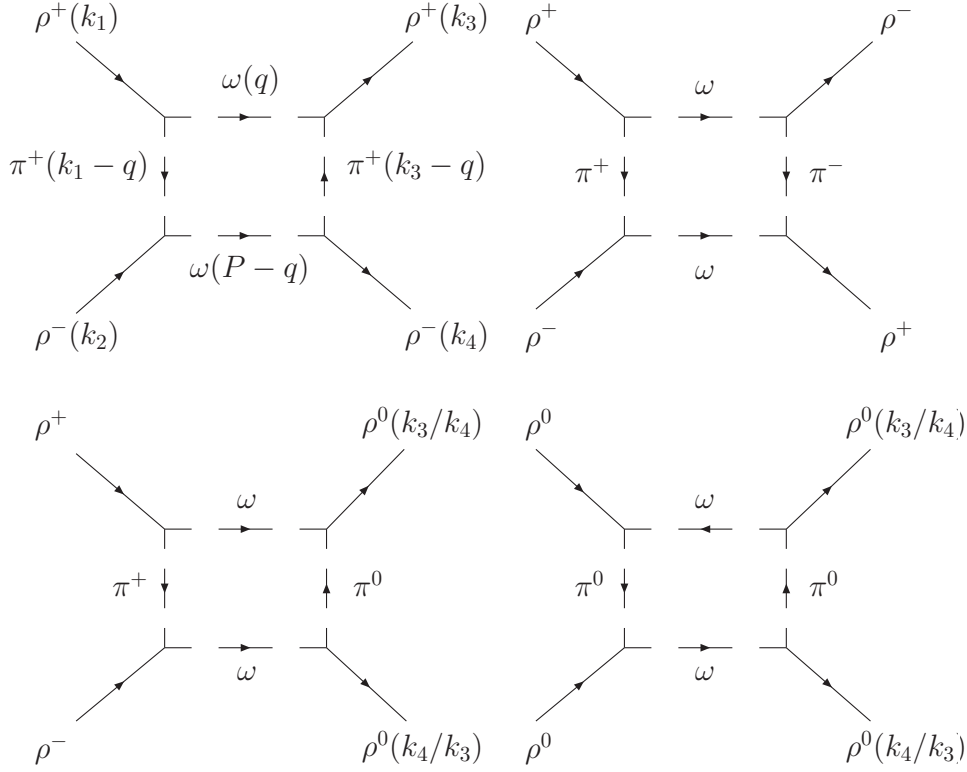}
\end{center}
\caption{ Anomalous-box diagrams for the two omega intermediate state.}
\label{fig:boxanomal}
\end{figure}

It is also interesting to evaluate the contribution of intermediate $\omega\omega$
state with anomaluous couplings, given by Fig. \ref{fig:boxanomal}.
The coupling $\rho\omega\pi$, with the renormalization that we use, can be found in \cite{hidekoroca} and is given by
\begin{equation}
{\cal L}_{VVP}=\frac{G'}{\sqrt{2}}\eps^{\mu\nu\alpha\beta}\langle \partial_\mu V_\nu \partial_\alpha V_\beta \,P\rangle
\end{equation}
with
\begin{center}
$ G'=\frac{3 {g'}^{2}}{4 \pi^2 f}$\hspace{0.6cm} $g'=-\frac{G_V M_\rho}{\sqrt{2}f^2}$\ ,
\end{center}
where $G_V\simeq 55\,MeV$ and $f_\pi=93\,MeV$. Thus, the vertex $\rho^+\pi^+\omega$ give
\begin{equation}
-it=iG'\eps^{\mu\nu\alpha\beta} \, q_\mu\, k_{1,\alpha}\,\eps_{\nu}(\omega)\,\eps_{\beta}(\rho^+)\ .
\end{equation}
At this point we use again the assumption that $\vec{k}_{i,j}\simeq 0$ which forces the index $\alpha=0$, and one obtain 
\begin{equation}
-it=iG'M_\rho\,\eps_{ijk} \,q_i \,\eps_{j}(\omega)\,\eps_{k}(\rho^+)\ .
\end{equation}
The amplitude corresponding to the first diagram of Fig. \ref{fig:boxanomal} is given by
\begin{eqnarray}
-it^{(\omega\omega)}&=&\int\frac{d^4 q}{(2\pi)^4}\,M^4_\rho\, {G'}^4\,\eps_{i_1j_1k_1}\,q_{i_1}\,\eps_{j_1}(\omega)\,\eps_{k_1}(\rho^+_1)\,\,
\eps_{i_2j_2k_2}\,q_{i_2}\,\eps_{j_2}(\omega)\,\eps_{k_2}(\rho^-_2)\nonumber\\&\times&\eps_{i_3j_3k_3}q_{i_3}\,\eps_{j_3}(\omega)\,\eps_{k_3}(\rho^-_4)\,\,\eps_{i_4j_4k_4}\,q_{i_4}\,\eps_{j_4}(\omega)\,\eps_{k_4}(\rho^+_3) \,\,\frac{1}{q^2-M^2_\omega+i\eps}\nonumber\\&\times&\frac{1}{(P-q)^2-M^2_\omega+i\eps}\,\frac{1}{(k_1-q)^2-m^2_\pi+i\eps}\,\frac{1}{(k_3-q)^2-m^2_\pi+i\eps}\ ,
\label{eq:anomal1}
\end{eqnarray}
which upon sum over the internal $\omega$ polarizations and simplifications done before,  leads to
\begin{eqnarray}
t^{(\omega\omega)}&=&-f(\eps_i)\,\frac{1}{15}\,M^4_\rho\,{G'}^4\,\int \frac{d^3 q}{(2\pi)^3}\, \vec{q}\,^4\,(-\omega^3_\pi+{k^0_3}^2\omega_\omega-4\,\omega^2_\pi\omega_\omega-4\,\omega_\pi\omega^2_\omega-\omega^3_\omega)\nonumber\\&\times&\frac{1}{(k^0_1+\omega_\omega+\omega_\pi)^2}\,\frac{1}{(\omega_\omega+\omega_\pi-k^0_1-i\eps)}\,\frac{1}{(\omega_\omega+\omega_\pi-k^0_3-i\eps)}\nonumber\\&\times&\frac{1}{\omega^3_\pi}\,\frac{1}{(P^0-2\omega_\omega+i\eps)}\,\frac{1}{(P^0+2\omega_\omega)}\,\frac{1}{\omega_\omega}\ ,
\label{eq:boxanomal1}
\end{eqnarray}
where $f(\eps_i)=6\,(\vec{\eps}_1\cdot \vec{\eps}_3) (\vec{\eps_2}\cdot\vec{ \eps}_4)+(\vec{\eps}_1\cdot \vec{\eps}_2)(\vec{\eps}_3\cdot \vec{\eps}_4)+(\vec{\eps}_1\cdot \vec{\eps}_4)(\vec{\eps}_2\cdot \vec{\eps}_3)$. When we evaluate the $\rho\rho$ interaction in $I=0$, the only one in this case, $f(\eps_i)$ is changed to 
\begin{equation}
f"(\eps_i)=7\,(\vec{\eps}_1\cdot \vec{\eps}_3) (\vec{\eps_2}\cdot\vec{ \eps}_4)+7\,(\vec{\eps}_1\cdot \vec{\eps}_4)(\vec{\eps}_2\cdot \vec{\eps}_3)+2(\vec{\eps}_1\cdot \vec{\eps}_2)(\vec{\eps}_3\cdot \vec{\eps}_4)\ ,
\end{equation}
which allows the projection over $S=0$ and $S=2$, and we finally obtain
\begin{eqnarray}
\tilde{V}^{(\omega\omega)}&=&-\frac{1}{30\,\pi^2}\,M^4_\rho\,{G'}^4\,\int_0^{q_{max}}dq\, \vec{q}\,^4\,(-\omega^3_\pi+{k^0_3}^2\omega_\omega-4\,\omega^2_\pi\omega_\omega-4\,\omega_\pi\omega^2_\omega-\omega^3_\omega)\nonumber\\&\times&\frac{1}{(k^0_1+\omega_\omega+\omega_\pi)^2}\,\frac{1}{(k^0_1+\frac{\Gamma}{4}-\omega_\omega-\omega_\pi+i\eps)}\,\frac{1}{(k^0_3-\frac{\Gamma}{4}-\omega_\omega-\omega_\pi+i\eps)}\nonumber\\&\times&\frac{1}{\omega^3_\pi}\,\frac{1}{(P^0-2\omega_\omega+i\eps)}\,\frac{1}{(P^0+2\omega_\omega)}\,\frac{1}{\omega_\omega}
\label{eq:boxanomal2}
\end{eqnarray}
and 
\begin{eqnarray}
t^{(\omega\omega,I=0,S=0)}&=&30\,\tilde{V}^{(\omega\omega)}\nonumber\\
t^{(\omega\omega,I=0,S=2)}&=&21\,\tilde{V}^{(\omega\omega)}\ .
\end{eqnarray}

\begin{figure}
\includegraphics[width=17cm]{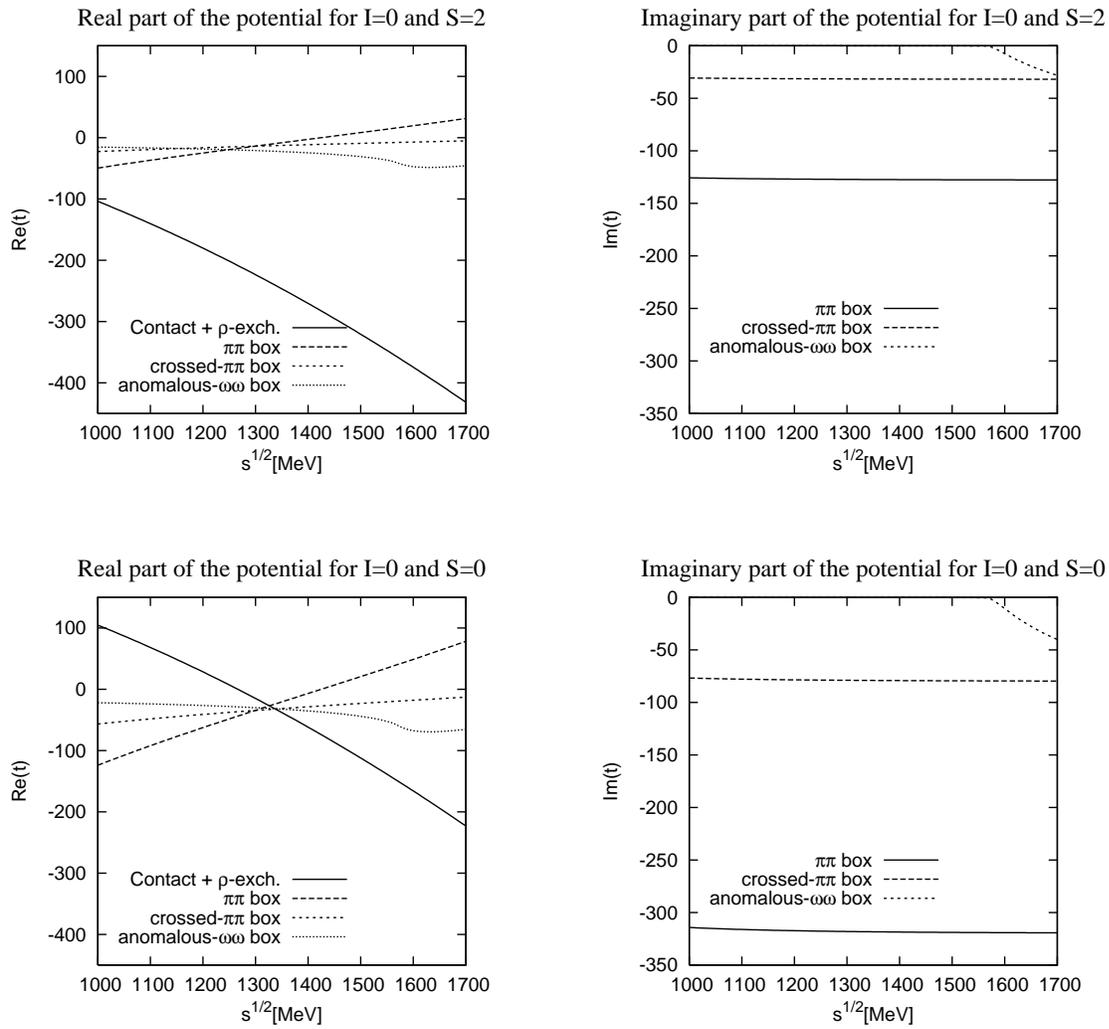}
\caption{Comparison of the real and imaginary parts of the different potentials for $I=0,\,S=2$ and $I=0,\,S=0$. }
\label{fig:potential}
\end{figure}
 
In Fig. \ref{fig:potential} we show the different contributions to the potential that we have
considered for $I=0,\,S=0$ and $I=0,\,S=2$. In the evaluation of the integrals we have taken the same cut off of $q_{max}=1200\,MeV$. For the sake of simplicity, the calculations are done without form factors. Their consideration does not change the conclusions that follow. Concerning the real parts, in the case of $S=2$ we observe that the most important contribution is the potential coming from the contact term and the $\rho$-exchange, which is very large and attractive, whereas the other terms are practically zero. For $S=0$ we observe that the individual contributions of the $\pi\pi$-box diagram, crossed-$\pi\pi$-box diagram and $\omega\omega$ term are comparatively larger with respect to the contact plus $\rho$-exchange term than in the case of $S=2$. Yet, we find a quite good cancellation of the $\pi\pi$ box plus crossed-$\pi\pi$ box and anomalous-$\omega\omega$ box terms, and the interaction is dominated by the contact plus $\rho$-exchange terms. However, the relatively larger contribution of the subdominant terms indicates that we should admit larger uncertainties in the position of the $f_0(1370)$ state than in the $f_2(1270)$ one. For the imaginary parts we see that the term of the $\pi\pi$ box, which allows for the decay of $\rho\rho$ in $\pi\pi$, is considerably larger than the others, and we obtain that the crossed-$\pi\pi$ box (decay in $4\pi$) only accounts for the $20\%$ of the $\pi\pi$ box, whereas the anomalous-$\omega\omega$ box is zero in our region of interest.

As we see, the crossed terms are reasonably smaller than the direct ones. We use this fact to omit the calculation of the crossed pion terms corresponding to the convolution of the two $\rho's$. This convolution implicitly includes the contribution of four intermediate pions when the two meson decay each into two pions. We could make one of two pions from one $\rho$ to be reabsorbed by the other $\rho$. We saw that the convolution of the $\rho's$ gave rise to a moderate width compared with the direct $\pi\pi$ box. Since the crossed-$\pi\pi$ box gives a smaller contribution than the direct term, we can expect the same to occur with the crossed terms from the convolution, giving rise to a small correction to a width which is already much smaller than the one obtained from the $2\pi$ decay. For this reason we omit the evaluation of these terms.

\section{Results with $V^{(\pi\pi)}$}

In view of the results obtained in the former section, here we show the results obtained considering only the contact term plus the $\rho$-exchange term and the imaginary part of the direct $\pi\pi$-box diagram. The $20\%$ extra contribution to the imaginary part of the $\pi\pi$-crossed box term is small compared with uncertainties in the width stemming from the use of the form factor of eq. (\ref{formfactor}).

\begin{figure}
\includegraphics[width=16.2cm]{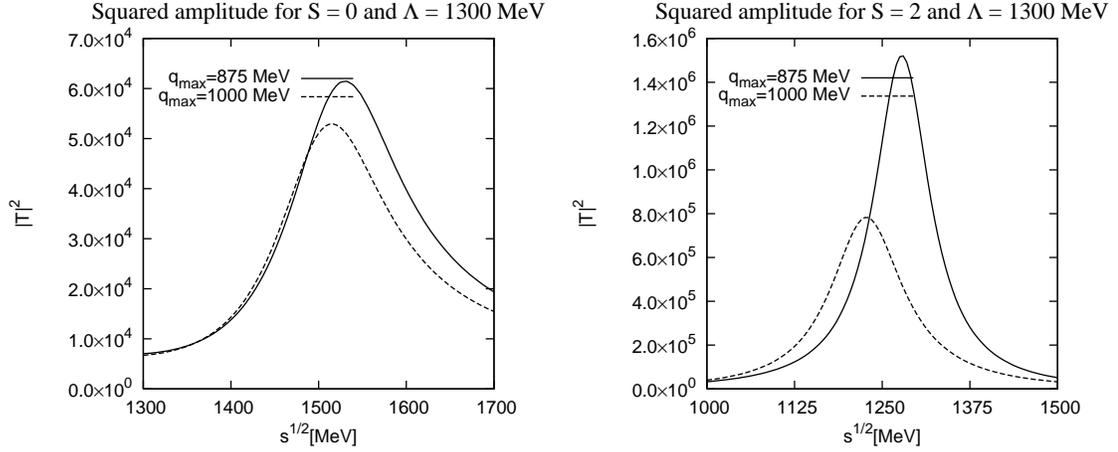}
\caption{$\vert T\vert^2$ taking into account the $\pi\pi$ box with
$\Lambda=1300\,MeV$, $q_{max}=875,1000\,MeV$ for $S=0$ and $S=2$. }
\label{fig:box1}
\end{figure}

In fig. \ref{fig:box1} we show the results for $\vert T\vert^2$ including the 
$\pi\pi$ box mechanism for a chosen value of $\Lambda=1300\,MeV$ and the two 
values of the cut off. What we observe is that the peak positions are barely 
changed with respect to fig. \ref{fig:convo}, however, the widths are now 
considerably larger. For the case of the $S=0$ state the width is of the order 
of $\Gamma=200\,MeV$, while for the case of the $S=2$ state is of the order of 
$110\,MeV$. The experimental situation is the following. The $f_2(1270)$ has a 
width of $\Gamma=185\,MeV$ mostly ($85$\%) coming from $\pi\pi$ decay \cite{pdg}. 
This means $\Gamma_{\pi\pi}\simeq 156\,MeV$, which should be considered in 
fair agreement with our results. For the case of the $f_0(1370)$ the width is 
$200-500\,MeV$ according to the PDG \cite{pdg}, with about $50$\% of the 
experiments providing a width around $200\,MeV$ in agreement with our 
findings. One might wonder whether the scalar state that we obtain, which has a
mass around $1500~MeV$, could not correspond to the  $f_0(1500)$. However, its
width of about $100 ~MeV$, out of which only 35\% comes from $\pi \pi $ decay
\cite{pdg}, clearly excludes it from being associated to the state that we have 
obtained around $1500~MeV$. 
On the other
hand, preliminary data from the Belle collaboration suggest that the peak of the
mass of the $f_0(1370)$ appears rather around $1470 ~MeV$ \cite{uehara}, which would agree 
better with our
findings. Yet, one should also take into consideration the thorough study of \cite{Bugg}, making a strong claim in favor of the $f_0(1370)$ with a mass around the nominal one of the PDG. Incidentally, this latter analysis relies upon a dispersive term dominated by $\rho\rho$ components. 

Our model provides some $4\pi$ decay coming from the $\pi \pi$ decay
of each $\rho$, which has been taken into account by means of the convolution
with the $\rho$ mass distribution. Also the crossed-$\pi\pi$ box diagram discussed in Section 9 gives rise to $4\pi$ decay. However, this cannot be the sole contribution
of $4\pi$ decay. In a recent paper  \cite{albaladejo}, extra decay channels of the type
of $\sigma \sigma$ are also considered which would increase the total width.\footnote{This work is being further extended and a more detailed discussion at
this point is untimely, but one should keep track of further
developments along this line to implement further improvements in our
approach.}

\begin{figure}
\includegraphics[width=17cm]{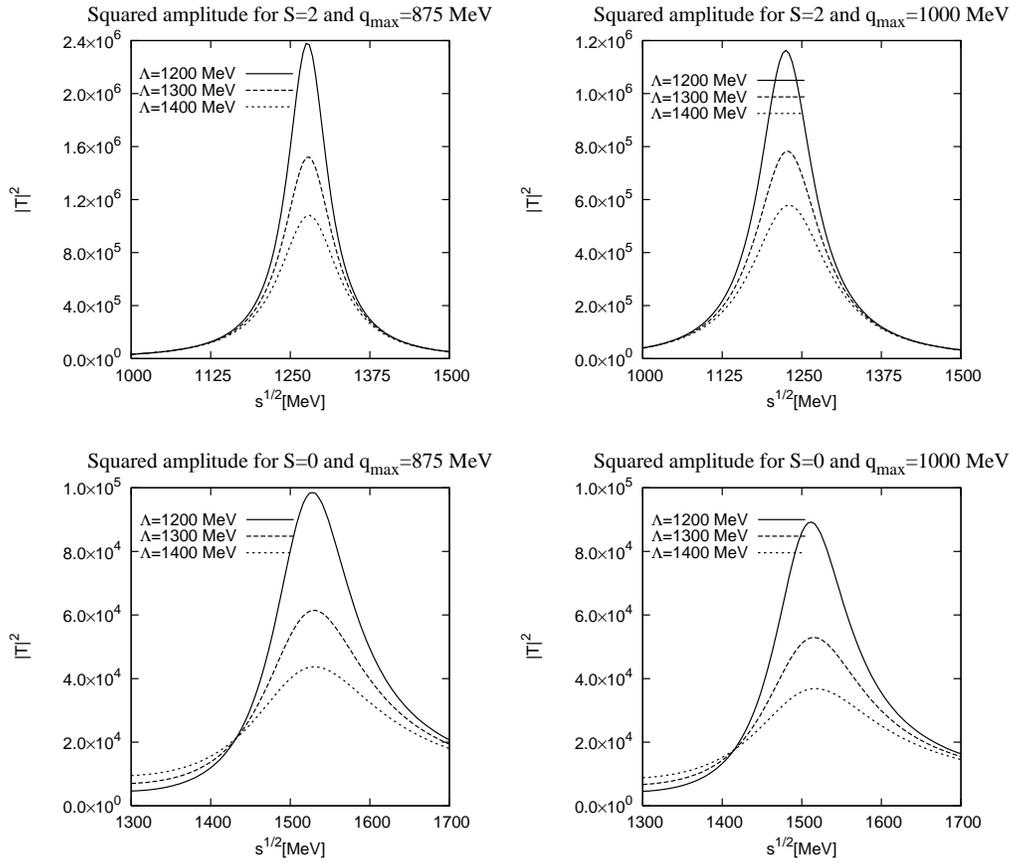}
\caption{$\vert T\vert^2$ taking into account the $\pi\pi$ box for different values of  $\Lambda=1200,1300, 1400\,MeV$ and $q_{max}=875,1000\,MeV$ for $S=0$ and $S=2$ }
\label{fig:box2}
\end{figure}

In order to show the sensitivity of the results to the meson decay form factor 
we show in fig. \ref{fig:box2} the results for different values of $\Lambda$, in the 
range of values used in \cite{titov1},\cite{titov2}. We take $\Lambda=1200, 1300$ 
and $1400\,MeV$. We can see that as $\Lambda$ grows, the width becomes larger 
but the peak position does not change. The dispersion on the values of the 
width gives us an indication of the theoretical uncertainties in this value. 
Yet, within these uncertainties in the position and the 
width, one can claim a reasonable agreement with data for these two states 
providing a big support for the idea of the two states as being dynamically 
generated from the $\rho\rho$ interaction given by the hidden gauge formalism.

\section{Conclusions}

We have made a study of the $\rho\rho$ interaction using the hidden gauge
formalism. The interaction comes from contact terms plus $\rho$ meson exchange
in the $t$-channel. Amongst all spin and isospin allowed channels in $s$-wave, we
found strong attraction, enough to bind the system, in $I=0,S=0$ and $I=0,S=2$.
We also found that in the case of $I=0,S=2$ the interaction was more attractive,
leading to a tensor state more bound than the scalar. The consideration of the
$\rho$ mass distribution gives a width to the two states, very small in the
case of the tensor state because of its large binding. However, the biggest
source of width comes from the decay into $\pi\pi$ that we have also studied
within the same formalism. We found the width much larger for the case of the
scalar state. We also studied the effect of the crossed-$\pi\pi$-box diagrams and the contribution of $\omega\omega$-intermediate states with anomalous couplings, which were found to play a minor role. The states obtained could be associated with the $f_0(1370)$ and
$f_2(1270)$, for which we found a qualitative agreement on the mass and
width. The findings of the paper give support to the idea that these
two resonances are dynamically generated from the $\rho\rho$ interaction, or in
other words, that they qualify largely as $\rho\rho$ molecules.

\section*{Acknowledgments}  

We would like to thank L. S. Geng for carefully checking the paper and the 
results and for useful comments. We also acknowledge useful information provided by J. A. 0ller and D. V. Bugg.
This work is partly supported by DGICYT contract number
FIS2006-03438. D. N. acknowledges partial support from the Austrian Science 
fund (FWF) under contract M979-N16.
This research is  part of
the EU Integrated Infrastructure Initiative Hadron Physics Project
under  contract number RII3-CT-2004-506078.

\end{document}